\documentclass[a4paper]{JHEP3}
\pdfoutput=1

\usepackage{graphicx}

\usepackage{amssymb,amsmath,amsfonts,amsbsy}

\newcommand{\setC}{\ensuremath{\mathbb{C}}}

\newcommand{\setZ}{\ensuremath{\mathbb{Z}}}
\newcommand{\setQ}{\ensuremath{\mathbb{Q}}}
\newcommand{\setR}{\ensuremath{\mathbb{R}}}

\newcommand{\TeV}{{\rm TeV}}

\newcommand{\abs}[1]{\ensuremath{\left| #1 \right|}}
\newcommand{\di}{\mathrm{d}}
\newcommand{\AdS}{\ensuremath{\mathrm{AdS}}}

\abstract{We study $M$--theory solutions involving compact hyperbolic
  spaces. The combination of a gap \emph{\`a la} Randall--Sundrum and
  the topology of an internal Riemann surface allows a geometrical
  solution to the hierarchy problem that does not require light
  Kaluza--Klein modes. We comment on the consequences of such a
  compactification for \textsc{lhc} physics. }

\author{Domenico Orlando{$^a$} and Seong Chan Park{$^b$}\\
Institute for the Physics and Mathematics of the Universe (IPMU)\\
The University of Tokyo
5-1-5 Kashiwa-no-Ha, Kashiwa City, Chiba 277-8583, Japan\\
Emails:  ${}^a$\email{domenico.orlando@ipmu.jp}, {${}^b$}\email{seongchan.park@ipmu.jp} 
}
\title{Compact hyperbolic extra dimension:\\
 a $M$--theory solution and its implications for the LHC}

\begin{document}

\section{Introduction}

An intriguing possibility is that we live in higher dimensional
spacetime which contains more than three spatial dimensions. Indeed,
the only candidate theories of quantum gravity are consistently
constructed in $D=10$ or $11$ dimensions, thus requiring six or seven
extra dimensions. Even though no direct evidence of extra dimension
has been found so far, several pressing problems in particle physics
including the big hierarchy problem~\cite{ArkaniHamed:1998rs,
  Antoniadis:1998ig, Randall:1999ee}, little hierarchy
problem~\cite{ArkaniHamed:2002qy}, flavor hierarchy problem, neutrino
mass problem, dark matter problem, supersymmetry breaking
problem~\cite{Randall:1998uk}, electroweak symmetry
breaking~\cite{ArkaniHamed:2001nc, Csaki:2002ur, Csaki:2003zu} and
several others have been already addressed in the context of higher
dimensions. (See \emph{e.g.}~\cite{Csaki:2004ay} for a recent review).

Now the question is what the extra dimensions actually look like, how
they behave and how they are stabilized. Because of lack of
experimental data, theorists have considered simple model geometries
like torii, spheres, (anti) de Sitter space, Calabi-Yau spaces,
etc\dots and studied their physical consequences. In this paper we
consider compact hyperbolic spaces which have been less intensively
studied in physics so
far~\cite{Kehagias:2000dg,Gauntlett:2000ng,Kaloper:2000jb,Chen:2003dca,Neupane:2003cs,Orlando:2007fv,Orlando:2006cc,Silverstein:2007ac,Douglas:2010rt,Greene:2010ch}
but enjoy several interesting mathematical properties (see
\cite{Farkas:1980,Thurston:1978aa} for a mathematical
introduction). In particular, there are two properties that are of
interest for our purposes:
\begin{enumerate}
\item A compact hyperbolic space comes equipped with two length
  scales: $\ell_c $ and $\ell_\Gamma$. The former, $\ell_c$, is
  related to local properties, such as the curvature and is fixed by
  the equations of motion (which are local in nature); the latter,
  $\ell_\Gamma$, is related to global properties, such as the volume,
  and is not affected by the equations of motion but appears in the
  expression of the effective Planck mass.
\item The volume (and the effective Planck mass) grows exponentially
  with the ratio of the two scales $\ell_\Gamma / \ell_c$, thus allowing 
  a more natural solution to the hierarchy problem.
\end{enumerate}
The combination of these length scales and a Randall--Sundrum gap
allows  a solution to the hierarchy problem that does not require
light Kaluza--Klein modes.

\bigskip

This paper is organized as follows. In Section~\ref{sec:string}, we
first report the M-theory solutions which contain a compact hyperbolic
space (\textsc{chs}) of the form $\AdS_{7-d}\times H_{d}/\Gamma \times
S^4$ and discuss the dimensional reduction, Kaluza-Klein spectrum and
the stability of the \textsc{chs}. In Section~\ref{sec:pheno}, we
examine the potentially realistic solutions containing at least four
non-compact dimensions, \emph{i.e.} $\AdS_{5,4}\times
H_{2,3}/\Gamma\times S^4$ and discuss a phenomenologically interesting
large volume compactification which brings the fundamental scale of
gravity down to the energies within the reach of the \textsc{lhc}. As
one of the most interesting predictions of large volume \textsc{chs},
which we call a \emph{Swiss Cheese Universe}, we examine the
possible production of microscopic black holes at the \textsc{lhc} and
its future upgrade with larger energy. Differently from other cases
like \textsc{rs}, \textsc{add} or \textsc{ued} models, \textsc{chs} it
is not necessarily accompanied by low scale Kaluza-Klein excitations
of higher dimensional fields even though its volume is large. Summary
and directions for future works are presented in the last section. In
the appendix, we collect mathematical facts regarding \textsc{chs}
which are extensively used in the text.

\section{M-theory solutions}
\label{sec:string}

\subsection{Solutions with maximally symmetric spaces}
\label{sec:solut-with-maxim}

The bosonic part of the eleven-dimensional supergravity action is
given by~\cite{Cremmer:1978km}
\begin{equation}
  S = \frac{M_{11}^2}{2} \int d^{11} x \:
  \sqrt{-G} \left( R - \frac{1}{2} \abs{F}^2 \right) -
  \frac{M_{11}^2}{12 } \int A \land F \land F,
\end{equation}
where $A$ is the three-form gauge field and $F=\mathrm{d}A$ its
associated field strength.  This leads to the following equations of
motion:
\begin{gather}
  \label{sugra}
  R_{MN} = \frac{1}{12} F^2_{MN} - \frac{1}{6}  G_{MN} \abs{F}^2 \, ,  \\
  \mathrm{d}*F + \frac{1}{2} F\wedge F = 0\ .  \label{4form}
\end{gather}
We will be interested in solutions which are direct products of
Einstein spaces  $\mathcal{M}_{11} = \mathcal{M}^{(0)} \times {\cal
  M}^{(1)} \times \mathcal{M}^{(2)} \times \cdots $, where ${\cal
  M}^{(0)}$ contains the time coordinate.  The Ricci tensor is in this
case block-diagonal, with each block proportional to the corresponding
metric:
\begin{equation}
  R_{MN}^{(a)} = k_a \, G_{MN}^{(a)} \quad \text{for } a = 0,1,2,\ldots
\end{equation}
An ansatz for the four-form field strength, which solves equation
(\ref{4form}) and preserves all geometric symmetries of the space,
is
\begin{equation}
  \label{flux}
  F = \sum_{A}Q_A\,  \omega^{(A)} \quad \text{with }
  \mathcal{M}^{(A)}\cap \mathcal{M}^{(B)} \not= \emptyset\,  .
\end{equation}
Here the $\mathcal{M}^{(A)}$ are four-dimensional sub-products inside
$\mathcal{M}^{(0)} \times \mathcal{M}^{(1)} \times \mathcal{M}^{(2)}
\times \cdots $ , and $ \omega^{(A)}$ is the corresponding volume
4-form.  Note that more than one flux components are only allowed when
$\mathcal{M}_{11}$ contains several factors of dimension lower than
four.  If $\mathcal{M}^{(A)}$ is compact, we must furthermore demand
the flux-quantization condition
\begin{equation}
  \int_{\mathcal{M}^{(A)}} \hskip -1mm F \ =\   \frac{2\pi n_A}{T_2}\  = 2\kappa^2 T_5 \, n_A \ ,
\end{equation}
where $T_2 = (2\pi^2/\kappa^2)^{1/3}$ is the fundamental-membrane
tension, $T_5$ is the tension of the M-theory fivebrane, and the $n_A$
are integers.

The Einstein equations \eqref{sugra} reduce, with the above product
ansatz, to the following set of algebraic conditions:
\begin{equation}
  \label{einseq}
  k_a + R \,  =\,   \frac{1}{2} \sum_{A\ni a}  \epsilon_{A} 
  Q_{A}^{\ 2}    \quad \text{for }  a = 0, 1, 2, \ldots,
\end{equation}
where $R = \frac{1}{6}\abs{F}^2 = \sum_a d_a k_a\ $\ is the total
Ricci scalar, $\epsilon_A= +$ or $-$ according to whether
${\mathcal{M}}^{(A)}$ is spacelike or timelike, and $d_a$ is the
dimension of the factor $\mathcal{M}^{(a)}$.  The sum on the
right-hand side of \eqref{einseq} runs over all sub-products that
contain $\mathcal{M}^{(a)}$.

Things simplify considerably when $F$ has only one non-vanishing
component, in which case $\mathcal{M}_{11}= \mathcal{M}_{4}\times
\mathcal{M}_{7}$, where $F= Q \times (\text{volume of }
\mathcal{M}_4)$.  While $\mathcal{M}_{4}$ and $\mathcal{M}_{7}$ could
be a priori products of simpler factors, it follows from equation
\eqref{sugra} that they must themselves be Einstein manifolds
(\emph{viz.} $R_{ij} = k \, g_{ij}$). Slightly abusing notation, we let
$k_4$ and $k_7$ be the corresponding Ricci scalars.  A simple
calculation gives
\begin{align}
   k_4 &=  \frac{Q^2}{3} & \text{and}& & k_7 &= - \frac{Q^2}{6} \, ,
\end{align}
for $\mathcal{M}_7$ Lorentzian, while for $\mathcal{M}_7$ Euclidean
the signs in the above relations must be reversed.  The two
best-studied solutions of this type~\cite{Freund:1980xh, Duff:1983gq}
are the near-horizon geometries of the membrane and of the five-brane:
$\AdS_{4} \times S^7$ and $\AdS_{7} \times S^4$. The fact that the
equations of motion only depend on the Ricci tensor implies that $S^4$
and $S^7$ can be replaced by products of spheres, with fixed ratios of
radii as imposed by the Einstein condition.  Likewise, instead of
$\AdS_4$ and $\AdS_7$ we may consider products of lower-dimensional
$\AdS$ times hyperbolic-space factors. Explicitly, $\AdS_4$ can be
replaced by $\AdS_2\times S^2$, and $\AdS_7$ by one of the following
possibilities: $\AdS_5\times H_2$, $\AdS_4\times H_3$, $\AdS_3\times
H_4$, $\AdS_3\times H_2\times H_2$, $\AdS_2\times H_5$ or
$\AdS_2\times H_3\times H_2$.  It is furthermore possible to mod out
the hyperbolic factors by a group of freely-acting isometries, $H_n\to
H_n/\Gamma$, so as to render them compact.  It is on this type of
vacua that we will focus our attention. We would like to stress once
more the fact that the curvatures of the subspaces are related. In
particular, for a geometry of the type $\AdS_{7-d} \times H_{d}
\times S^4$, the Ricci scalars satisfy
\begin{equation}
  \frac{R[ \AdS_{7-d1} ]}{7-d} = \frac{R [H_{d}] }{d} = - \frac{R
    [S^4] }{8} = - \frac{1}{\ell_c^2} \, ,  
\end{equation}
where we introduced the length scale $\ell_c$.  It is convenient to
express the volumes of the compact spaces in terms of this length
scale and -- for the hyperbolic part -- a second scale $\ell_\Gamma$ (for
details see Appendix~\ref{sec:comp-hyperb-space}). For $d = 2,3$, the volumes are
\begin{align}
  V[S^4] &= \frac{9 \pi^2}{8} \ell_c^{4} \, , &
  V[H_2 / \Gamma_g] &\sim \pi \ell_c^2 e^{\ell_\Gamma / \ell_c} \, ,
  \\ &&
  V[H_3 / \Gamma_n] &\sim \pi \sqrt{2} \ell_c^3 e^{\sqrt{2}\ell_\Gamma / \ell_c} \, ,   
\end{align}
where $\Gamma_g$ is a lattice such that $H_2 / \Gamma_g $ is a
genus-$g$ Riemann surface and $\Gamma_n = PSL(2, \mathcal{O}_n)$ as
in Appendix~\ref{sec:comp-hyperb-space}.

These solutions are in general not protected by supersymmetry; this
implies in particular that we should care about their stability. In
our analyses we will deal with the breathing modes of the compact
$H_{d}$ and $S^4$ internal manifolds which, in an effective action
description, are to be represented by scalar fields. The stability
(with respect to small fluctuations) will then translate into the
positivity of the squared mass for such fields, a condition that can
be relaxed into satisfying a Breitenlohner--Freedman bound when the
space--time is of the Anti-de-Sitter type.

\subsection{Dimensional reduction}
\label{sec:dangerous-modes}

In this section, we want to discuss the stability of the
$\AdS_{7-d} \times H_{d} / \Gamma \times S^4$ solutions we found
above. In particular we show that, in a consistent way, we can limit
ourselves to the study of two scalar modes, corresponding to the
volumes of the two compact internal manifolds. The strategy is the
following: first we reduce from eleven to seven dimensions on the
four-sphere with flux, and then we perform a geometric reduction
on the hyperbolic manifold.

\bigskip

To perform the first step (from eleven to seven dimensions), we need
to consider the bosonic part of the seven-dimensional supergravity
action~\cite{Pernici:1984xx,Duff:1982yg,Townsend:1983kk}. We will use
the same notation as in~\cite{Pernici:1984xx}. By construction, mass
term contributions can only come from the scalars $T_{ij}$ or the
5-plet of antisymmetric tensors $S_{\alpha \beta \gamma,I}$.
\begin{itemize}
\item The scalar modes are collected in the coset $SL(5,
  \setR)/SO(5)$, hence we can choose a single representative, namely
  $\det (T_{ij})$ which corresponds to \emph{the volume of the
    four-sphere}.
\item The global $SO(5)$ symmetry allows us to consider only one of
  the (spacetime) three-forms $S$. In this case, a Chern-Simons mass
  term could arise and be in principle negative. The relevant equation
  of motion reads
  \begin{equation}
    \di S = \lambda * S \, ,
  \end{equation}
  where $\lambda$ is a symplectic Majorana spinor. Taking the Hodge
  dual and differentiating, one gets a Klein-Gordon equation for the
  topological mass term:
  \begin{equation}
    \left( \triangle + (-)^{mr+1} \lambda^2 \right) S = 0 \, ,
  \end{equation}
  where $\triangle $ is the Laplacian, $m=7$ is the spacetime
  dimension and $r=3$ is the degree of $S$. One can verify that the
  mass contribution is positive and does not lead to instabilities.
\end{itemize}

\bigskip

The second step consists in going from seven to four dimensions. In
this case, there are no gauge fields and the compactification is
completely geometric. This means that Kaluza-Klein instabilities can
only come from scalar modes, and more precisely from the zero mode
since on a compact manifold the Laplace operator is always
non-negative. In other words, the only potentially dangerous mode is
\emph{the overall volume of the compact manifold} $H_{d}/
\Gamma$. The analysis of the two modes we have identified is the
object of the next section.

\subsection{Stability of the breathing modes}
\label{sec:stab-breath-modes}

Consider a metric ansatz obtained as a warped product with three factors:
\begin{equation}
  \di s^2 = \di s^2_{M_0 }(x) + \mathrm{e}^{2 \varphi_1(x)} \di s^2_{M_1}
  +  \mathrm{e}^{2 \varphi_2(x)} \di s^2_{M_2} \, ,
\end{equation}
where $M_0, M_1, M_2 $, have respectively dimension $d_0, d_1, d_2$,
$M_1$ and $M_2$ are compact (we will refer to them as internal), and the
$\varphi_i(x)$ only depend on the coordinates in $M_0$. We wish to
study an action of the type
\begin{equation}
  S = \frac{M_{11}^2}{2}\int \di^{11} x \: \sqrt{ g^{(11)}} \left(
    R^{(11)} - V (x) \right) \, ,
\end{equation}
where $R^{(11)}$ is the Ricci scalar in $11$
dimensions and $V$ is a potential that depends on $\varphi_i(x)$.

Integrating out the internal coordinates one obtains an effective
action in $d_0$ dimensions, which has to be brought to the usual
Einstein-Hilbert form via a Weyl rescaling:
\begin{equation}
  g_{\mu \nu}^{(d_0)} = \exp \left[ 2 \frac{d_1 \varphi_1 + d_2 \varphi_2}{d-2} \right] g_{\mu \nu}^{(d_0)} (M_0) \, .
\end{equation}
The contributions from the curvatures of the internal manifolds are
collected into an effective potential $\bar V( \varphi_1,
\varphi_2)$. Moreover, the scalars need to be rescaled to get the
canonical dynamical term. The resulting action reads:
  \begin{equation}
    S = \frac{M_{d_0}^2}{2 } \int \di^{d_0} x \: \sqrt{- g^{(d_0)}} \left[
      R^{(d_0)} - \frac{1}{2} \partial_\mu \Phi_1 \partial^\mu \Phi_1 -
      \frac{1}{2} \partial_\mu \Phi_2 \partial^\mu \Phi_2 -
      \bar V (\Phi_1, \Phi_2) \right] \, ,
  \end{equation}
where
\begin{equation}
\label{eq:bar-potential}
  \bar V (\varphi_1, \varphi_2) =  e^{-2 \left( d_1 \varphi_1 + d_2 \varphi_2 \right) /\left( d_0 - 2
    \right)} \left( - e^{-2\varphi_1(x)} R^{(1)} - e^{-2\varphi_2(x)}
    R^{(2)} + V (\varphi_1, \varphi_2) \right) \, ,
\end{equation}
and
\begin{equation}
  \begin{cases}
    \Phi_1 = \sqrt{\frac{18}{\left(d_1 + d_2 \right)
        \left( d - 2 \right)}} \left( d_1 \varphi_1 + d_2 \varphi_2
    \right) \\
    \Phi_2 = \sqrt{\frac{2 d_1 d_2}{d_1+d_2}} \left( \varphi_1 -
      \varphi_2 \right) \, .
  \end{cases}
\end{equation}
Note that $\Phi_1$ describes the overall volume of the compact part,
and $\Phi_2$ describes a mode where one manifold grows while the other
shrinks.

As we have stressed already, the type of backgrounds we obtain after
compactification are $\AdS$, which means that tachyons can be
accepted if they don't cross the Breitenlohner-Freedman bound.  It is
convenient to rewrite this condition in terms of the effective
potential $V(\Phi_1, \Phi_2)$ and its value around the stationary
points. Taking into account the contribution from the negative
curvature, one obtains the stability condition:
\begin{equation}
  \label{eq:FB-bound}
  - \frac{d_0-1}{4 \left( d_0 - 2 \right)} \left\langle \bar V \right\rangle + m_i^2 \ge 0 \, ,
\end{equation}
where $\left\langle V\right\rangle$ is the value of the potential at the critical points $\Phi_i = \bar \Phi_i$ and $m_i^2$ are the eigenvalues of the Hessian matrix
\begin{equation}
  H_{ij} = \left. \frac{ \partial^2 \bar V}{\partial \Phi_i \partial \Phi_j } \right|_{\Phi_k = \bar \Phi_k} \, .
\end{equation}

We are now in position to treat the $M$-theory backgrounds of the
form $\AdS_{7-d} \times H_{d} / \Gamma \times S^4$, where the
hyperbolic space has been divided out by a discrete isometry group
$\Gamma$ which makes it compact. The potential $V(\varphi_1,
\varphi_2)$ is due to the presence of a four-form field on the $S^4$
part,
\begin{equation}
  V ( \varphi_1, \varphi_2 ) = \frac{Q^2}{2} e^{- 8 \varphi_2} \, .
\end{equation}
The effective potential in Eq.~\eqref{eq:bar-potential} becomes:
\begin{equation}
  \bar V ( \varphi_1, \varphi_2 ) = e^{-2 \left( d \varphi_1 -4 \varphi_2 \right)/\left(5-d \right)} \left( - \frac{d}{2} e^{-2 \varphi_1} - 2 e^{\varphi_2} + \frac{Q^2}{2} e^{ - 8 \varphi_2} \right),
\end{equation}
and one finds that for any value of $d$ the solution we found above is
a minimum and thus stable without having to invoke any \textsc{bf}
arguments.

\section{Phenomenological Implications}
\label{sec:pheno}

\subsection{4D effective action, the hierarchy problem and the Swiss Cheese Universe}

Among the M-theory backgrounds of the form $\AdS_{7-d} \times H_{d}/\Gamma \times S^4$, two cases are  particularly interesting from the phenomenological point of view as they contain four or more non-compact dimensions:  
\begin{itemize}
\item{case i:} $\AdS_{5} \times H_{2}/\Gamma \times S^4$,
\item{case ii:} $\AdS_{4} \times H_{3}/\Gamma \times S^4$.
\end{itemize}
Of these two cases, the former is more promising as the flat 4D space
(visible universe) can be embedded into $\AdS_{5}$ just like in
Randall-Sundrum models \cite{Randall:1999ee, Randall:1999vf}.  For the latter case to be realistic,
the $\AdS_{4}$ space would have to be deformed to a nearly flat space
by some (yet unknown) mechanism.

For a given background, we can get the 4D effective action after
dimensional reduction, as discussed in the previous section.  First of
all, the resulting effective action contains the conventional
Einstein-Hilbert term,

\begin{equation}
S_4 = \int \di^4 x \sqrt{g^{(4)}} \frac{M_4^2}{2}R^{(4)} + \dots \, ,
\end{equation}

where the Planck scale in 4D, $M_4=1/\sqrt{8\pi G_N} = 2.435 \times10^{18}$ GeV,  is related to the fundamental gravity scale in eleven dimensions, $M_{11}$,  as

\begin{equation}
  M_4^2 = M_{11}^9 \ell_c^7 {\mathcal V}_{\Gamma}.
\end{equation}

Here $\ell_c$ denotes the curvature radius of compact spaces and
${\mathcal V}_\Gamma$ the dimensionless volume of the \textsc{chs}. As
explained in Appendix~\ref{sec:comp-hyperb-space},
$\mathcal{V}_\Gamma$ depends only on the topology via the scale
$\ell_\Gamma$:
\begin{equation}
{\mathcal V}_{\Gamma}= \begin{cases}
  \frac{9 \pi^3}{8} e^{\ell_\Gamma / \ell_c} & \text{for $H_2/\Gamma$,} \\
    \frac{9\sqrt{2} \pi^3}{8} e^{\sqrt{2} \ell_\Gamma / \ell_c} & \text{for $H_3 / \Gamma$.}
  \end{cases}
\end{equation}

Since ${\mathcal V}_\Gamma$ is not bounded from above, in principle,
$M_{11}$ can be \emph{arbitrarily small}. On the other hand, for both
compact manifolds the first eigenvalue of the Laplacian is fixed by
$\ell_c$ as follows:
\begin{align}
  \Delta_1[H_2 / \Gamma] &= \frac{1}{4 \ell_c^2} \, ,&
  \Delta_1[H_3 / \Gamma] &= \frac{3}{8 \ell_c^2} \, , &
  \Delta_1[S^4] &= \frac{2}{3 \ell_c^2} \, .
\end{align}
This means that $\ell_c^{-1}$ must be big enough to explain why the
extra dimensions have not been experimentally accessible yet. More
precisely, since no Kaluza--Klein (\textsc{kk}) excitation of the
graviton has been observed, the lightest \textsc{kk} state, of order
$\ell_c^{-1}$, must be heavier than 1 TeV (the details
depending on the scenario). A particularly interesting situation is
realized when the gravity scale is fairly close to the electroweak
scale which is the only energy scale entering in the standard model as
an input\footnote{\textsc{qcd} scale may be regarded as another scale but
  it is dynamically realized by strong interactions.}. For instance,
$M_{11}\sim 1$ TeV can be obtained provided that $\ell_c^{-1}
\sim 1$ TeV and
\begin{equation}
  \mathcal{V}_\Gamma \sim
  M_4^2 {/\TeV^2}\sim 10^{30} \Rightarrow \ell_\Gamma =
  \begin{cases}
    \sim 66{/\TeV} & \text{for $H_2 / \Gamma $,} \\
    \sim 46{/\TeV} & \text{for $H_3 / \Gamma $.}
  \end{cases}
\end{equation}
In this case the big hierarchy between Planck scale and electroweak
scale is understood as a consequence of the topology of the
\textsc{chs}.\footnote{See also \cite{Kaloper:2000jb} where a similar discussion has
  been carried out in a completely phenomenological way. In this
  sense, this paper fills a theoretical gap.} Since (in the two
dimensional case) the genus of the surface grows exponentially with
the ratio $\ell_\Gamma / \ell_c$ (as shown in
Appendix~\ref{sec:comp-hyperb-space}), one can say that the large
number of ``holes'' in the extra dimension is responsible for the
large hierarchy. In other words, gravity is weak because we live in a
``\emph{Swiss Cheese Universe}''.

There are two different ways of embedding flat four dimensional space
in $\AdS_5$ \emph{\`a la} Randall--Sundrum: \textsc{rs}-1 \cite{Randall:1999ee} and
\textsc{rs}-2 \cite{Randall:1999vf}. In particular, in an \textsc{rs}-1 construction, the
presence of a large warp factor provides a different explanation for
the hierarchy, related to the exponential red-shift in scales. In
\textsc{rs}-1, two branes are introduced at the boundaries ($y=0 $ and
$y=\ell_5$) of a slice of $\AdS_5$ described by the line element $ds^2
= e^{-y/\ell_c}\eta_{\mu\nu}dx^\mu dx^\nu + dy^2$. Since the energy
scale in the warped space depends on the location as
$\Lambda(y)=e^{-y/(2\ell_c)}M_{UV}$, the brane located at $y=0$ is
called UV-brane and the other brane at $y=\ell_5$ is called
IR-brane. Defining $M_{UV}=M_5$ (the five-dimensional gravity scale),
the IR scale is exponentially suppressed as $M_{IR} = M_5 \,
e^{-\ell_5/(2\ell_c)}$. Here $M_5$ is further related to $M_{11}$ by
the relation $M_5^3 = M_{11}^9 \ell_c^6 \mathcal{V}_\Gamma$, so that
\begin{equation}
  M_{IR} = M_5 e^{-\ell_5/(2\ell_c)}
\sim 3.26 \   M_{11}^3 \ell_c^2  \exp [\frac{2\ell_\Gamma - 3 \ell_5}{6 \ell_c}] \, .
\end{equation}
This result is interesting as \emph{``anarchy''}, $M_{IR}\sim
M_{11}\sim \ell_c^{-1}$, is realized when
\begin{equation}
  \boxed{2\ell_\Gamma \sim 3 \ell_5}  \, .
\end{equation}
In this case, the hierarchy between $M_5$ and $M_{IR}$ or $M_{11}$ 
originates from the large warp factor and the volume of the compact
hyperbolic space.  Using now the fact that $M_4 $ and $M_5$ are
related by dimensional reduction,
\begin{equation}
  M_4^2 = 2 M_5^3 \ell_c \, ,
\end{equation}
we can find the values of $\ell_5$ and $\ell_\Gamma$ that are
required to reproduce the desired energy scales:
\begin{equation}
  \frac{M_5}{M_{11}} \sim 10^{10} \sim \mathcal{V}_\Gamma^{1/3} \sim e^{\ell_5/(2\ell_c)}  \Rightarrow \ell_5 \sim 44{/\TeV} \, .
\end{equation}

\bigskip

Let us stop a moment and summarize what we have seen so far. The
compactification on $H_2 / \Gamma \times S^4$ brings two length
scales, $\ell_c$ and $\ell_\Gamma$. The former is related to the
eleven-dimensional Planck mass $M_{11} = \ell_c^{-1}$. Since it
appears in the expression for the Kaluza--Klein modes, $\ell_c$ is
fixed experimentally to be $\ell_c \ge 1{\TeV}$. The other scale,
$\ell_\Gamma$, governs the compactification volume and appears in the
expression for the five-dimensional Planck mass. In a \textsc{rs}-1 scenario,
we have yet another scale, $\ell_5$, which contributes to the IR
gravity scale in four dimensions. When the two scales are related by
$2\ell_\Gamma \sim 3 \ell_5$ an anarchy is realized, where the
eleven-dimensional Planck mass $M_{11}$ and the infrared gravity scale
$M_{IR}$ are both of the same order of magnitude as $\ell_c^{-1} \sim
1{\TeV}$. Requiring the four-dimensional Planck mass to be of
order $M_4 \sim 2.4\times 10^{15}{\TeV}$ fixes the values of the two length
scales to $\ell_\Gamma \sim 66{/\TeV}$ and $\ell_5 \sim
44{/\TeV}$, and gives a five-dimensional Planck mass of order
$M_5 \sim 1.4 \times 10^{10}{\TeV}$.

\bigskip

As a last comment we would like to point out that in a \textsc{rs}-2 scenario,
gravity is trapped at $y=0$ and no particular relation to the IR scale
is realized by the warped geometry. The hierarchy is obtained again by
$\ell_\Gamma \sim 66{/\TeV}$ as we have already seen.

\subsection{Implications for the LHC and beyond}

The low scale for $M_{11}$ in the Swiss Cheese Universe opens exciting
possibilities for experimental tests at the \textsc{lhc}. It is
instructive to compare our case with the well known extra dimensional
scenarios, e.g., \textsc{add}~\cite{ArkaniHamed:1998rs, Antoniadis:1998ig, ArkaniHamed:2002qy}, \textsc{rs}~\cite{Randall:1999ee, Randall:1999vf} and
\textsc{ued}~\cite{Antoniadis:1990ew, Appelquist:2000nn, Cheng:2002iz} (see also~\cite{Park:2009cs}) since some of the features are common while others
are quite distinctive.

\begin{itemize}
\item $M_G\sim 1{\TeV}$: Gravity becomes strong at low energy as
  in \textsc{add} and \textsc{rs}.  We do not have a big hierarchy
  problem. In \textsc{ued}, $M_5 \sim (M_4^2/R)^{1/3}\gg
  1{\TeV}$.
\item $M_{\textsc{kk}}\sim 1{\TeV}$: Differently from
  \textsc{add}, our case does not allow very light \textsc{kk} states
  thus gets less constrained by cosmological and astrophysical
  observations. This is rather similar to the \textsc{rs}-1 and
  \textsc{ued} cases.
\item $M_G\sim M_{\textsc{kk}}$: The effect of strong gravitation is not
  accompanied by \textsc{kk} gravitons. This feature is unique in our case
  since still at least a few \textsc{kk} gravitons or \textsc{kk} states of standard
  model fields can be seen in \textsc{rs}-1 or \textsc{ued}.
\item Microscopic black holes: Once we access the transplanckian
  region $\sqrt{s}\gg M_{11}$, a semiclassical description of the
  scattering process based on eleven dimensional (super)gravity theory
  becomes valid \cite{'tHooft:1987rb, Banks:1999gd}. In fact, the Schwarzschild radius for a colliding
  particles with the CM energy $\sqrt{s}$ in eleven dimension,
  \begin{equation}
    R_{\text{Sch}} = \left(\frac{8}{3}\frac{\sqrt{s}}{M_{11}}\right)^{1/8}\frac{1}{M_{11}},
  \end{equation}
  is much larger than the eleven dimensional Planck length
  $M_{11}^{-1}$. With the expected CM energy
  $\sqrt{s}=10(14)\TeV$ at the \textsc{lhc} and
  the even higher energy $\sqrt{s}=100 \TeV$ at
  the future upgraded collider (\textsc{vlhc}), we expect that eleven
  dimensional microscopic black holes can be produced if $M_{11}$ is
  sufficiently low or $\mathcal{V}_\Gamma$ is big enough \cite{Giddings:2001bu, Dimopoulos:2001hw}. The parton
  level production cross section $\hat{\sigma}$ is estimated using the
  Hoop conjecture, which is in good agreement with the numerical estimation \cite{Eardley:2002re, Yoshino:2002tx}:
  \begin{equation}
    \hat{\sigma} \simeq F_{11} \left(\frac{\sqrt{\hat{s}}}{M_4}\right)^{1/4} \frac{\mathcal{V}_\Gamma^{9/8}}{M_4^2}
    \simeq \frac{1.9 \times 10^{-33}}{1{\TeV^2}}
    \left(\frac{F_{11}}{45}\right)
    \left(\frac{\sqrt{\hat{s}}}{10{\TeV}}\right)^{1/4}
    \mathcal{V}_\Gamma^{9/8} \, ,
  \end{equation}
  where $F_{11}$ is the form factor\footnote{$F_{4+n}= \pi
    \left(\frac{2}{(1+(n+2)^2/4)^{1/(n+1)}}\right)^2
    \left(\frac{\Gamma(n+3/2)}{(n+2)\pi^{(n+3)/2}}\right)^{2/(n+1)}$
    is derived taking the angular momentum into account in $D=4+n$
    dimensions.} and $\sqrt{\hat{s}}$ is the CM energy in the parton
  frame (we assume $M_{11}\sim \ell_c^{-1}$).  Even though there is a
  large Planck energy suppression, in principle, the large volume of
  hyperbolic space can overcome it and the cross section can be
  sizable. In Figure~\ref{fig:sigma}, we plot the black hole
  production cross section by proton-proton collisions at the
  \textsc{lhc} ($\sqrt{s}=14{\TeV}$, solid) and V\textsc{lhc}
  ($\sqrt{s}=100{\TeV}$, dotted), respectively. We introduce a
  convenient parameter $\hat{V}$ which is defined by the relation $
  M_4 /\sqrt{\mathcal{V}} = 1{\TeV}/ \sqrt{\hat{V}}$ and consider
  the most interesting region $\sqrt{\hat{V}}\in [0.1, 10]$. The cross
  section grows fast as $\hat{V}$ becomes bigger since the parton
  level cross section is proportional to $\hat{V}^{9/8}$ and the scale
  of the strong gravity so as the threshold energy is lower with
  larger $\hat{V}$. The minimum mass is chosen to be of the order of
  $10\times M_4 /\sqrt{\mathcal{V}}$ in this calculation. Since the
  \textsc{kk} scale is high in \textsc{chs}, the dominant decay
  mechanism is Hawking radiation to the zero modes or the Standard
  Model particles. Grey-body factors for these particles have been
  developed in \cite{Ida:2002ez, Ida:2005ax,  Ida:2006tf, Harris:2005jx, Casals:2005sa, Casals:2006xp} and also Monte-Carlo event generators are available \cite{Dai:2007ki, Frost:2009cf}.
  \begin{figure}
    \centering
    \includegraphics[width=.9\columnwidth]{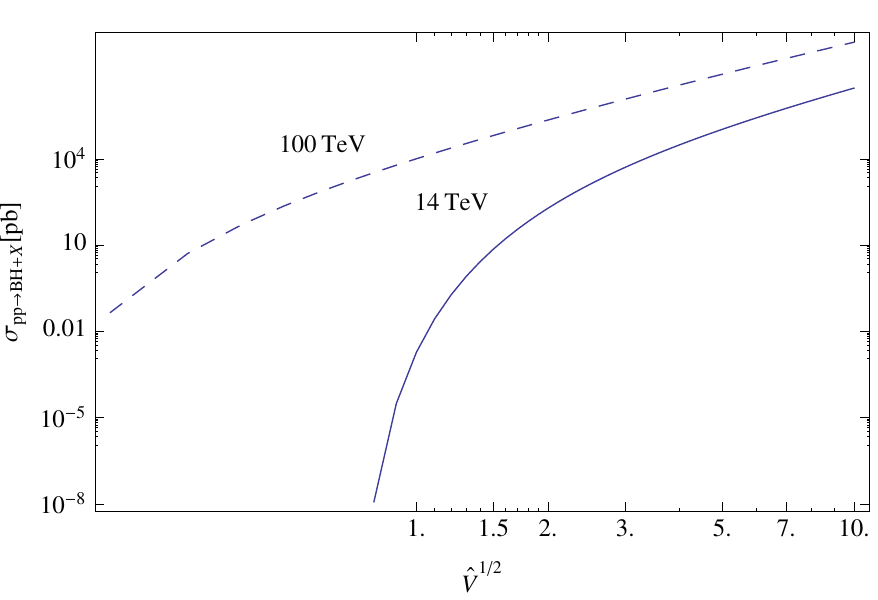}  
    \caption{Production cross-sections of black hole at the
      \textsc{lhc} ($\sqrt{s}=14{\TeV}$) and \textsc{vlhc}
      ($\sqrt{s}=100{\TeV}$).  }
    \label{fig:sigma}
  \end{figure}
\item String winding modes: Naively one would expect that a relatively
  high \textsc{kk} scale might imply the existence of low energy
  winding states. However, for a Riemann surface, the scale
  $\ell_\Gamma $ and the length of the shortest non-trivial geodesic
  $m$ are related to the volume $V$ by~\cite{Mumford:1971,Chang:1977}:
  \begin{equation}
    m \ell_\Gamma \sim V  \, .
  \end{equation}
  This means that the shortest path around the holes grows
  exponentially with $\ell_\Gamma / \ell_c$, like the volume
  itself. It follows that the energy of the winding modes is of order
  $E_{\text{wind}}\sim \tau \ell_c^2 / \ell_\Gamma \mathcal{V}$, where
  $\tau=1/l_s^2$ is the string tension, which is much bigger that the
  other scales we have considered so far.
\end{itemize}

\section{Discussion}

In this note, we have shown that $\AdS_5 \times H^2/\Gamma \times S^4$
is a solution to eleven dimensional $M$--theory. This leads to a gap,
$\ell_5$, in the fifth dimension of $\AdS_5$ \emph{\`a la}
Randall-Sundrum, and a large volume for compact hyperbolic space (fixed
by the topology). We can speculate that all the scales of physics,
namely (i) the Planck scale ($M_{\text{Planck}} = \frac{1}{\sqrt{8\pi G}}$),
(ii) a high scale which might be responsible for the coupling
unification ($M_{\text{GUT}} < M_{\text{Planck}}$) and (iii) the scale
of electroweak symmetry breaking ($M_{\text{EW}}$) are genuinely
realized via the geometrical properties of higher dimensions starting
from a single scale, the fundamental scale of eleven dimensional
supergravity $M_{11} \sim 1/\ell_c$ as follows:
\begin{align}
  M_{\text{Planck}} &= M_4 \sim M_{11}^4 \ell_c^3 e^{\ell_\Gamma/(2\ell_c)}, \\
  M_{\text{GUT}} &= M_5 \sim M_{11} e^{\ell_5/(2\ell_c)}, \\
  M_{\text{EW}} &= M_{11},
\end{align}
where the topological information on the \textsc{chs} is encoded in
$\ell_\Gamma$.

Since a large $\ell_\Gamma$ corresponds to a large genus for \textsc{chs},
we may call this spacetime the \emph{Swiss Cheese Universe}.

The Swiss Cheese Universe predicts low energy strong gravity but
without light Kaluza--Klein excitation or light stringy winding
states. Thus, the observation of microscopic black holes at high energy
colliders, even possibly at the \textsc{lhc}, without low scale exotic
Kaluza--Klein particles would be a clear signature of the Swiss Cheese
Universe which we believe should be examined in further detail.

\section{Acknowledgements}

We would like to thank Susanne~Reffert for dicussions. DO would like
to thank Costas~Bachas, Costas~Kounnas and Marios~Petropoulos for
collaboration on a related project.

This work is supported by the World Premier International Research
Center Initiative (WPI initiative) by MEXT. SCP is also supported by
the Grant-in-Aid for scientific research (Young Scientists (B)
21740172) from JSPS, Japan.

\appendix

\section{Compact hyperbolic space}
\label{sec:comp-hyperb-space}

An $n$-dimensional maximally symmetric (Einstein) space is a
pseudo-sphere in $n+1$ dimensions, \emph{i.e.} the locus of the points
in $\setR^{n+1}$ satisfying the equation:
\begin{equation}
  \epsilon_0 ( X^0 )^2 + ( X^1 )^2 + \dots + ( X^{n-1} )^2 + \epsilon_n ( X^n )^2 = \epsilon L^2 ,
\end{equation}
where the three $\epsilon$ parameters are signs. In particular,
$\epsilon_0$ and $\epsilon_n$ specify the signature of the embedding
$\setR^{n+1}$ space. The hyperbolic space $H_n $ is the Euclidean
manifold corresponding to the choice $\epsilon_0 = -1$, $\epsilon_n = 1$,
$\epsilon = -1 $ (other choices of the signs lead to $\AdS_n$,
$\mathrm{dS}_n$ or $S^n$).

By construction $H_n$ has (negative) constant curvature. In terms of
Riemannian geometry, the relevant tensors can be written as:
\begin{align}
  R_{abcd} &= -\frac{1}{L^2} \left( g_{ad} g_{bc} - g_{ac}
    g_{bd} \right) \, , \\
  Ric_{ab} &= R^{c}_{\phantom{c}acb} =
  -\frac{1}{L^2} \left( n - 1 \right) g_{ab} \, , \\
  R &= Ric^{a}_{\phantom{a}a} = -\frac{1}{L^2} n \left(n - 1
  \right) \, .
\end{align}
For later convenience we also introduce a length scale $\ell_c$ via
\begin{equation}
  \ell_c^2 = \frac{L^2}{n-1} = \frac{n}{\abs{R}} \, .
\end{equation}

Another very useful description can be given in terms of Lie groups. A
maximally symmetric space is identified with the coset $G/T$ (we
quotient with respect to the action of $T$ on the left, $g \sim g t$),
where $T$ is the maximal subgroup in the group $G$. In particular, one
can show that
\begin{equation}
  H_n = \frac{SO(1,n)}{SO(n)} \, .  
\end{equation}
The first obvious consequence is that $SO(1, n)$ is the group of
isometries of $H_n$.

A particularly convenient choice of coordinates, covering the whole
manifold, is given by the so-called Poincar\'e coordinates. The line
element takes the form
\begin{equation}
  \di s^2 = g_{ab} \di u^a \di u^b = \frac{L^2 }{(u^1)^2 } \left( (\di u^1)^2 + \dots + (\di u^n)^2 \right)  \, ,
\end{equation}
where $u^i \in \left( 0, + \infty \right)$. In such coordinates it is
evident that such manifolds have infinite volume, \emph{i.e.} the
integral $\left(\int \di u^1 \dots \di u^n \, \sqrt{g} \right)$
diverges.

\bigskip

Starting from $H_n$, it is possible to construct \emph{compact}
manifolds of constant negative curvature by taking the quotient with
respect to the action of a freely acting discrete group $\Gamma
\subset SO(1,n)$ (from now on we call this a lattice). It is worth
emphasizing that, although $H_n$ and $H_n / \Gamma$ share the same
local properties, the global properties are completely different. The
construction is general, in the sense that any closed manifold of
negative constant curvature can be written as a quotient $H_n /
\Gamma$.

One of the most important results in the study of these manifolds is
\emph{Mostow's rigidity theorem}~\cite{Mostow:1968aa}. It states that
the geometry of a finite volume hyperbolic manifold of dimension
greater than two is determined by its fundamental group\footnote{We
  are not considering the case of cusped hyperbolic three-manifolds or
  incomplete metrics.}. In particular, this means that once we have
fixed the curvature ($L$) and the volume, there are no more moduli.
The theorem is not valid in $d=2$ dimensions. In fact, a Riemann
surface of genus $g>1$ (which can always be represented as quotients
$H_2 / \Gamma$, $\Gamma \subset SO(1,2)$), has a $6 \left( g - 1
\right)$-dimensional moduli space.

In view of the following physical applications, we will now concentrate
on the lower-dimensional $d=2$ and $d=3$ examples.

\paragraph{$d=2$, or, Riemann surfaces.}

\begin{figure}
  \centering
  \begin{tabular}{cp{1em}cp{1em}c}
    \includegraphics[width=.3\columnwidth]{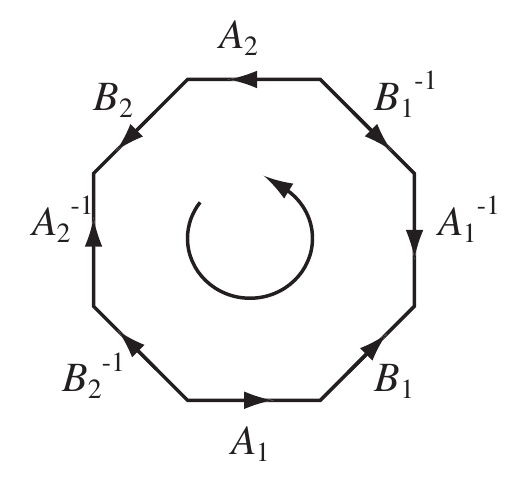}
    &&     \includegraphics[width=.25\columnwidth]{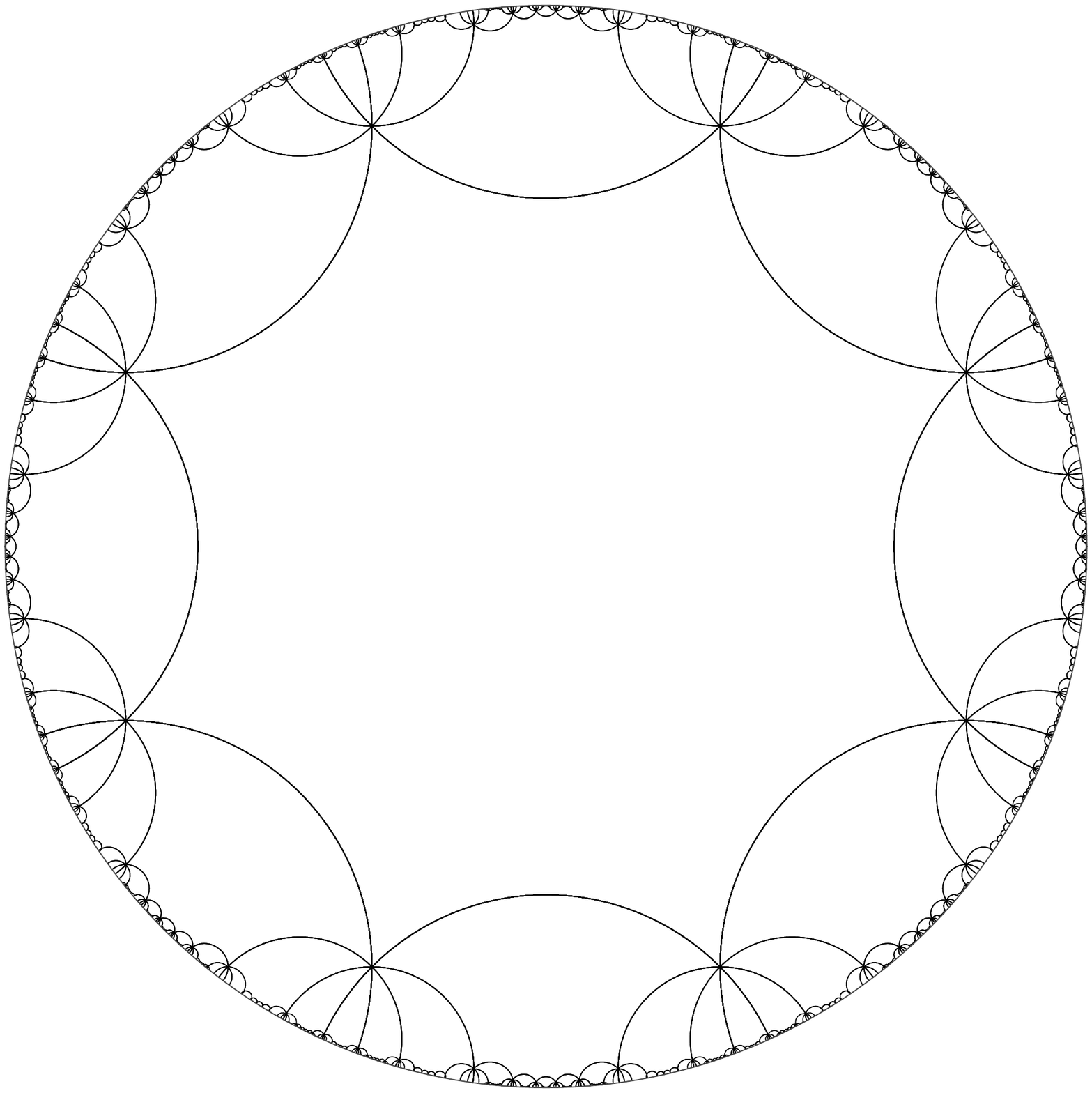}
    &&     \includegraphics[width=.3\columnwidth]{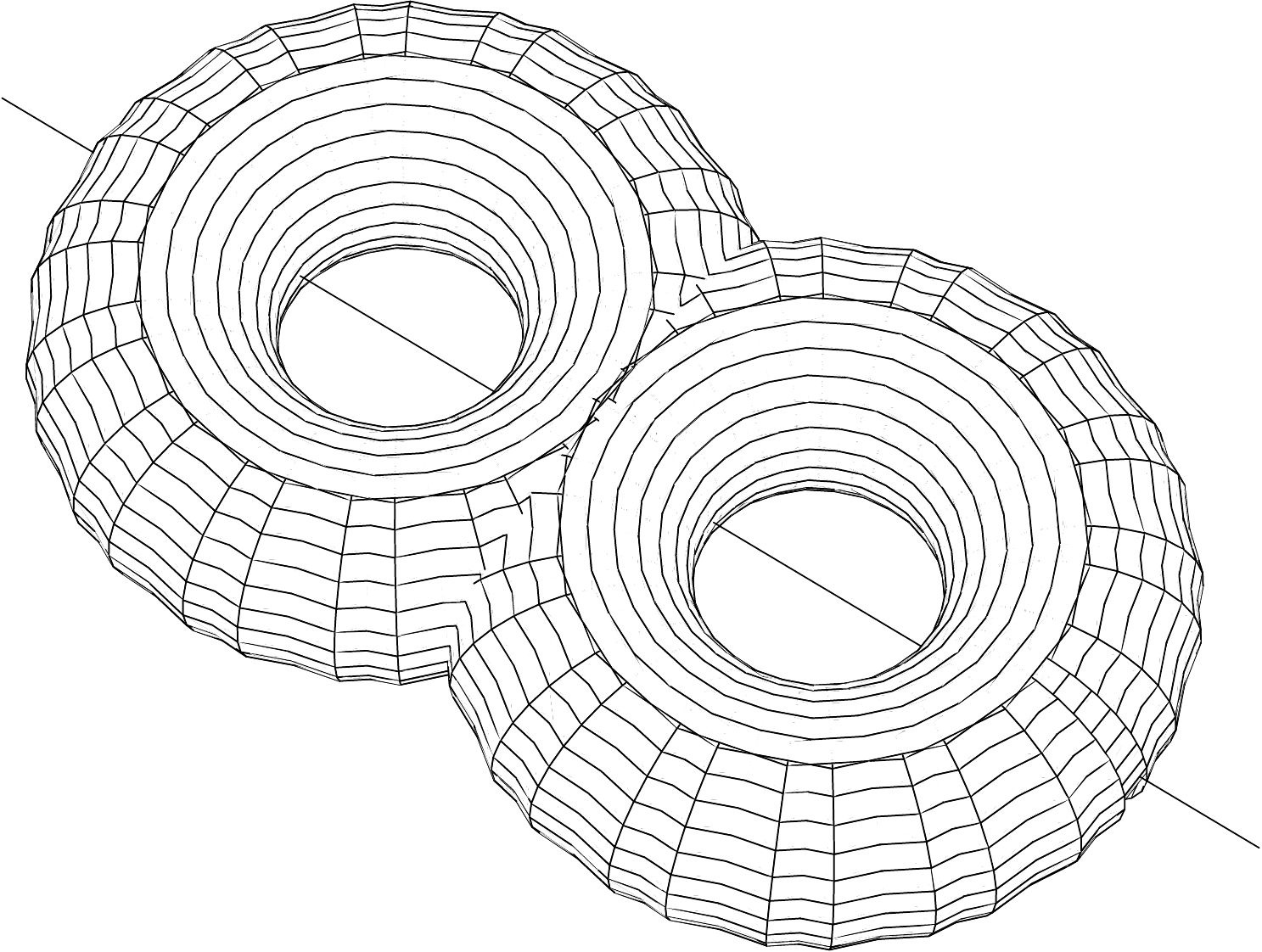} \\
    (a) && (b) && (c)
  \end{tabular}
  \caption{Different representations of a Riemann surface of genus
    $g=2$. (a) Standard fundamental polygon
    $A_1B_1A_1^{-1}B_1^{-1}A_2B_2A_2^{-1}B_2^{-1} $; (b) tessellation of
    the Poincar\'e disc; (c) embedding in $\setR^3$. }
  \label{fig:Genus-two}
\end{figure}

Depending on their genus, Riemann surfaces can be endowed with a
metric that can be spherical (genus $g=0$), flat (genus $g=1$) or
hyperbolic ($g\geq2$). \\
Any surface of genus $g$ can be described in terms of a polygon. In
particular we define the metric fundamental polygon as the $\left( 4 g
  + 2 \right)$-gon in which the edges are pairwise identified and the
standard fundamental polygon as the $\left(4g \right)$-gon in which
the edges are pairwise identified and all the vertices are identified.
In the usual notation, an $n$-gon is represented as a string of $n$
letters with exponent $\pm 1$ depending on the orientation of the edge
with respect to an arbitrary positive one. The same letter is used for
pairs of identified edges.  The most familiar example is the torus
that can be built out of a parallelogram lattice in $\setR^2$,
opposite sides being identified and having opposite orientation, that
is to say the standard fundamental polygon is
$ABA^{-1}B^{-1}$. Equivalently, the torus is also to be obtained by an
hexagonal lattice (metric fundamental polygon) with sides
$ABCA^{-1}B^{-1}C^{-1}$.\\
The fundamental polygons can be seen as elementary cells for a
tessellation of the hyperbolic plane $H_2$. In
Figure~\ref{fig:Genus-two} we show one such tessellation, corresponding
to a genus two surface, on the Poincar\'e disc.\\
There are two properties of Riemann surfaces that will be of interest
in the following: the volume, which is function of the genus $g$ and
the curvature
\begin{equation}
  V[H_2/ \Gamma_g] = 4 \pi L^2 \left( g - 1 \right) = 8 \pi \frac{1}{\abs{R}} \left( g - 1 \right) \, ,  
\end{equation}
and the fact that the scalar Laplacian is massive and the gap only
depends on the curvature.One can always choose a Riemann surface such
that:
\begin{equation}
  \Delta_1 \ge \frac{1}{4 L^2} = \frac{\abs{R}}{8}\, .
\end{equation}
Note that there are two independent parameters that fix these two
quantities: a ``local'' parameter $L$ and a ``global'' parameter
$g$. This is an important difference with respect to the case of the
sphere, where both volume and mass gap are fixed by the
curvature. This is the key property of compact hyperbolic manifolds
that allows us to propose a different mechanism for dealing with the
hierarchy problem.

It is convenient to introduce a scale parameter $\ell_\Gamma$ related to the
global properties of the surface, to express the volume. Recall that
the volume of a $d$--ball of radius $\ell_\Gamma$ in $H_d$ is given by
\begin{equation}
  V_d ( \ell_\Gamma ) = S_{d-1} L^d I_{d-1} ( \ell_\Gamma/L ) \, ,
\end{equation}
where $S_{d-1}$ is the surface area of a Euclidean $\left( d - 1
\right)$--sphere and $I_{d-1}(\ell_\Gamma / L)$ is the integral
\begin{equation}
  I_{d-1}(\ell_\Gamma/L) = \int_0^{\ell_\Gamma/L} \sinh(\xi)^{d-1} \di \xi \, . 
\end{equation}
We can then define the typical length $\ell_\Gamma$ of a Riemann surface of
genus $g$ via:
\begin{equation}
  4 \pi L^2 \left( g - 1 \right) \sim 2 \pi L^2 \cosh (\ell_\Gamma/L)  \Rightarrow
  l \sim L \log (g) \, , 
\end{equation}
and the volume is
\begin{equation}
  V[H_2/\Gamma] \sim \pi L^2 e^{\ell_\Gamma / L} \, .  
\end{equation}

\paragraph{Hyperbolic three-manifolds.}

The three-dimensional case is the first example in which Mostow's
theorem holds. In particular, this means that the geometry of a
compact hyperbolic three-manifold is fixed by the choice of the
lattice and hence by its volume. Here we show how this volume can be
computed for a large family of lattices by using algebraic techniques. \\
The hyperbolic space $H_3$ is the coset $SO(1,3)/SO(3)$. This implies
that its group of isometries is $SL(2, \setC) \sim SO(1,3)$ and in
particular any discrete lattice used to compactify is $\Gamma \subset
SL(2, \setC)$. A convenient choice of $\Gamma$, depending on an
integer number $d$ leads to the so--called Bianchi manifolds for which
it is possible to give explicit expressions for the volume and mass
gap. Let $n$ be a square-free integer (\emph{i.e.} $n$ does not
contain any perfect square as factor). Consider the quadratic field
$\setQ (\sqrt{-n}) = \{a+b \sqrt{-n} | a,b \in \setQ\}$, \emph{i.e.}
the two-dimensional vector space on $\setQ$ generated by $1$ and
$\sqrt{-n}$. Let $\mathcal{O}_n \subset Q(\sqrt{-n}) $ be the ring of
integers in this field. Explicitly:
\begin{equation}
  \mathcal{O}_n = \setZ [ \omega], \hspace{1em} \omega =
  \begin{cases}
    \frac{1}{2} \left( -1 + \sqrt{-n} \right) & \text{if $n=3 \mod 4$,} \\
    \sqrt{-n} & \text{otherwise.}
  \end{cases}
\end{equation}
(Note that when $n = 3 \mod 4$, $\omega $ is an algebraic integer
since it satisfies the equation $\omega^2 + \omega + \frac{\left( n + 1
  \right)}{4} = 0$).\\
The most simple example is $\setQ(\imath)$, the so-called the Gaussian
rationals, \emph{i.e.} the field $\setQ(\imath) = \{ a + \imath b |
  a,b \in \setQ \}$ and $\mathcal{O}_1 $ is the ring of Gaussian
integers $\mathcal{O}_1 = \setZ [\imath]= \{ a + \imath b | a,b \in
  \setZ \}$.\\
$\mathcal{O}_d$ is a lattice in $\setC$, so we can consider the
lattice $PSL(2, \mathcal{O}_n ) \subset PSL(2, \setC)$ and the
compactifications $H_3 / \Gamma$ where $\Gamma \subset PSL(2, \setC)$.
When one sees $\mathcal{O}_n$ as a lattice in $\setC$, its area is
given by $\mathrm{vol} ( \mathcal{O}_n ) = \sqrt{D}/2 $ where, $D$ is
the discriminant:
\begin{equation}
  D (\mathcal{O}_n ) =
  \begin{cases}
    n & \text{if $n=3 \mod 4$,} \\
    4n & \text{otherwise .}
  \end{cases}
\end{equation}
We need to take into account the difference between considering the
lattice on $\setC$ and on $PSL(2, \setC)$, and this is done by using
the formula~\cite{Sarnak:1983aa}
\begin{equation}
  \mathrm{vol} (H_3 / PSL(2, \mathcal{O}_n))  = \frac{ L(\chi_{n}, 2)}{24} L^3 D^{3/2} = \frac{ L(\chi_n, 2) }{3} \left( \frac{3D}{2\abs{R}} \right)^{3/2} \, ,
\end{equation}
where $L(\chi_{n},s) $ is the Dirichlet $L$-series for the principal
character $\chi_{n}$ (for our purposes it is worth to remark that $1
\le L(\chi_{n},2) \le \zeta(2) = \pi^2 / 6 \sim 1.644$).  
Also in this case, it is convenient to introduce a length scale $\ell_\Gamma$,
related to the volume via
\begin{equation}
  \frac{ L(\chi_{n}, 2)}{24} L^3 D^{3/2} \sim 4 \pi L^3 \left(
    \frac{\sinh(2 \ell_\Gamma / L )}{4} - \frac{\ell_\Gamma}{2 L }\right) 
  \Rightarrow l \sim \frac{3}{4} L \log (n) \, .
\end{equation}
The volume of the three-manifold is given by
\begin{equation}
  V[H_3 / \Gamma ] \sim \frac{\pi L^3}{2} e^{2 \ell_\Gamma / L} \, .  
\end{equation}

Using number--theoretical techniques, it is possible to evaluate the
mass gap. Just like in the Riemann surface case, this only depends on
the global parameter $L$ and is
\begin{equation}
  \Delta_1 \ge \frac{3}{4 L^2} = \frac{9 \abs{R}}{2} \, .
\end{equation}
Note that, just like before, volume and mass gap depend on two
parameters $L$ and $n$ that can be adjusted independently.

%\bibliography{chs}

\end{document}